\documentclass[reprint,twocolumn,longbibliography,english,amsmath,amssymb,aps]{revtex4-1}
\usepackage{graphicx}% Include figure files
\usepackage{pgfplots}
\usepackage{mathtools}
\usepackage{hyperref}
\usepackage{dcolumn}% Align table columns on decimal point
\usepackage{bm}% bold math
\usepackage{float}
\begin{document}

\preprint{APS/123-QED}

\title{Random matrix ensemble for the covariance matrix of Ornstein-Uhlenbeck processes with heterogeneous temperatures}

\author{Leonardo S. Ferreira and Fernando L. Metz}
\affiliation{Physics Institute, Federal University of Rio Grande do Sul, 91501-970 Porto Alegre, Brazil}
\author{Paolo Barucca}
\affiliation{
 Department of Computer Science, University College London, WC1E 6BT London, United Kingdom
}

\begin{abstract}
  We introduce a random matrix model for the stationary covariance of multivariate Ornstein-Uhlenbeck processes
  with heterogeneous temperatures, where the covariance is constrained by the Sylvester-Lyapunov equation.
  Using the replica method, we compute the spectral density
  of the equal-time covariance matrix characterizing the stationary states, demonstrating that this model undergoes
  a transition between stable and unstable states. In the stable regime, the
  spectral density has a finite and positive support, whereas negative eigenvalues emerge in the unstable regime.
  We determine the critical line separating these regimes and show
  that the spectral density exhibits a power-law tail at marginal stability, with an exponent independent
  of the temperature distribution.
  Additionally, we compute the spectral density of the
  lagged covariance matrix characterizing the stationary states of linear transformations of the original
  dynamical variables. Our random-matrix model is potentially interesting to understand the spectral properties of empirical correlation matrices appearing
  in the study of complex systems. 
\end{abstract}

\maketitle

\section{Introduction}

The history and success of random matrix theory demonstrate that diverse data sets, regardless of their origin, can display
universal patterns in their spectral decomposition \cite{Tracy2009}. Random matrix theory has been applied to various fields, including quantum
physics \cite{wigner1958,MIRLIN2000259}, climate time series \cite{Santos2020}, functional MRI \cite{kwapien2000temporal} and financial
data \cite{plerou1999universal,Laloux1999}.
The underlying interactions in large complex systems -- such as neural networks, ecosystems, and stock
markets \cite{Song2011,Sandoval2012,Munnix2012} -- are typically inferred from the empirical covariances among the system's states.
Random matrix theory plays a pivotal role in the problem of estimating the covariance matrix from noisy data obtained from
high-dimensional dynamical systems \cite{ledoit2004well,Bun2017,burda2022cleaning}.

The dynamical variables describing complex systems
evolve in time according to
nonlinear differential equations \cite{Metz2024}.
The stability of fixed-points is crucial to understand the behavior of these systems.
For example, a stable ecosystem is often associated with a stable
fixed-point \cite{may1972will,allesina2012stability}, where
species abundances exhibit small fluctuations around their average values \cite{krumbeck2021fluctuation,Marti2018}. However, as model parameters change, stable fixed-points can become
unstable \cite{Neri2020,allesina2012stability,Patil2024}, leading to a variety of
dynamical behaviors, including chaos and periodic oscillations \cite{Sompolinsky1988,Bunin2017,Marti2018,Metz2024}.
A popular conjecture suggests that complex systems have adapted to operate at marginal stability \cite{Beggs2003,Munoz2018,Moran2019,Bouchaud2024}, i.e., near
the critical point of the stability transition. In this situation, large fluctuations of the dynamical variables result in power-law distributions
of various quantities.
For instance, economic models show power-law fluctuations in prices and firm sizes \cite{Moran2019}, models of ecosystems
display a power-law singularity in their power-spectrum \cite{krumbeck2021fluctuation}, and neuronal activity distributions exhibit power-law tails \cite{Munoz2018}.
Despite the importance of the covariance matrix as an experimentally accessible quantity, a systematic analysis of how its spectral properties
behave across the stability transition has remained elusive.

Random matrix theory provides benchmark models for empirical covariance
matrices \cite{Laloux1999}, allowing to
distinguish between eigenvalues representing pure randomness from those reflecting genuine correlations
within the system \cite{Plerou2002,Noh2000,Bun2017}. Random-matrix models of covariance matrices rely on {\it ad hoc} assumptions for
the statistics of the matrix entries \cite{Laloux1999,burda2004free,Burda2004,burda2005spectral,Noh2000,burda2010random,barucca2020eigenvalue}. The most
prominent example is the so-called Wishart ensemble \cite{Wishart1928}, where the covariance matrix is constructed by multiplying a pair of
rectangular matrices with Gaussian-distributed entries. The main advantage of the Wishart ensemble is that many spectral
properties can be analytically computed \cite{marchenko1967distribution,Katzav2010,Majumdar2012,Bun2017,Castillo2018}. However, a
significant drawback is that this model does not account for the interactions within the system, making it
unsuitable for studying how the stability transition affects the spectral properties of the covariance matrix.

In general, there is no direct relationship between the covariance matrix
and the interactions in complex systems
due to the nonlinearity of the dynamics. However, if the
dynamical equations can be linearized near the stability transition \cite{Kwon2005}, the dynamics
simplify to a multivariate Ornstein-Uhlenbeck process (MVOU) \cite{risken1996fokker,meucci2009review}. In this linear regime, the stationary covariance matrix
satisfies the Sylvester-Lyapunov equation \cite{godreche2018characterising,fyodorov2023fluctuation}, which depends solely on the diffusion and interaction
matrices defining the process. This framework opens the possibility of introducing covariance matrix ensembles that
explicitly account for interactions, potentially improving our understanding of stability transitions through
the analysis of the covariance spectrum.

Building on previous work \cite{barucca2014localization}, we introduce an ensemble of covariance matrices derived 
from the stationary states of reversible MVOU processes, where each random matrix instance is a solution of the Sylvester-Lyapunov equation. Thus, this
ensemble incorporates the constraints
imposed by the Sylvester-Lyapunov equation, resulting in a model where the covariance matrix is determined by the interaction and diffusion matrices.
By considering fully-connected Gaussian interactions and a diagonal diffusion matrix, with elements representing the local temperatures of the dynamical
variables, we study how the temperature distribution influences the spectrum of the covariance across the stability transition.

Using the replica method of disordered systems \cite{NishimoriBook}, we derive analytic results for the spectral densities of both the equal-time covariance matrix
and the lagged covariance matrix, with the latter resulting from a generic linear transformation of the original dynamical process. These results
enable a systematic investigation of how the stability transition impacts the spectral density. In the stable regime, the eigenvalues
are positive and the spectral density $\rho_{S}(\lambda)$ of the equal-time covariance $\boldsymbol{S}$ is supported on a finite interval, while the unstable phase is marked by the appearance
of negative eigenvalues. At marginal stability, all eigenvalues remain positive
and $\rho_{S}(\lambda)$ exhibits a power-law tail $\rho_{S}(\lambda) \propto \lambda^{-5/2}$, with
an exponent independent of the temperature distribution.
This finding suggests that the power-law decay of $\rho_{S}(\lambda)$ is an universal property of complex systems at marginal stability.

The paper is organized as follows. In section \ref{secMVOU}, we introduce reversible MVOU processes and derive general solutions for the covariance and
lagged covariance matrices as functions of coupling strengths and temperatures. Section \ref{RMT} defines the specific ensembles of covariance matrices.
In Section \ref{specdensity}, we present analytic results
for the spectral densities of the covariance matrices for arbitrary temperature
distributions. Section \ref{secresults} provides numerical results for specific temperature distributions, confirming
our theoretical predictions and illustrating the behavior of the spectral density of the equal-time covariance across the stability transition.
Finally, the last section summarizes our results and indicates possible directions of future research.
The paper includes an appendix detailing the replica calculations of the spectral densities.

%%%%%%%%%%%%%%%%%%%%%%%%%%%%%%%%%%%%%%%%%%%%%%%%%%%%%%%%%%%%%%%%%%%%%%%%%%%%%%%%%%%%%%
%%%%%%%%%%%%%%%%%%%%%%%%%%%%%%%%%%%%%%%%%%%%%%%%%%%%%%%%%%%%%%%%%%%%%%%%%%%%%%%%%%%%%%%

\section{Multivariate Ornstein-Uhlenbeck processes}
\label{secMVOU}

We consider $N$ dynamical variables $X_1(t),\dots,X_N(t)$ that may represent
the abundances of different species in an ecosystem, the neuronal activities in the brain, or the stock prices
of a financial system. The degrees of freedom $\boldsymbol{X}(t) = (X_1(t),\dots,X_N(t))^{T}$ evolve in time according
to a  multivariate Ornstein-Uhlenbeck process (MVOU), which is represented
by the coupled system of stochastic differential equations
\begin{equation}
\label{MVOU}
d\boldsymbol{X} = -\boldsymbol{A}\boldsymbol{X}dt + \boldsymbol{\eta}\sqrt{dt},
\end{equation}
where $\eta_i(t)$ is a Gaussian noise with zero mean and covariance
\begin{equation}
\langle \eta_i(t) \eta_j(t^{\prime}) \rangle_{\eta} = 2D_{ij} \delta(t - t^{\prime}). 
\end{equation}  
The element $A_{ij}$ of matrix $\boldsymbol{A}$
quantifies the influence of  $X_j(t)$ on $X_i(t)$,
while the element $D_{ij}$ of  the symmetric positive-definite
matrix $\boldsymbol{D}$ controls the covariance
between $\eta_i(t)$ and $\eta_j(t)$. The coupling matrix $\boldsymbol{A}$ and the diffusion matrix $\boldsymbol{D}$ completely specify the model. The
Gaussian noise variables $\eta_1(t),\dots,\eta_N(t)$ account for environmental random perturbations.

The central object of our interest is the two-time covariance matrix $\boldsymbol{S}(s,t)$ of the dynamical variables. The elements of this $N \times N$ matrix are given by
\begin{equation}
S_{ij}(s,t) = \langle X_i(s) X_j(t) \rangle_{\eta},
\end{equation}  
where $\langle \dots \rangle_{\eta}$ denotes the ensemble average over the Gaussian noise.
Equation (\ref{MVOU}) converges to a stable stationary solution provided the real parts of all eigenvalues of $\boldsymbol{A}$ are positive \cite{risken1996fokker}. In this case, the
joint distribution of $X_1(t),\dots,X_N(t)$ evolves to a multivariate Gaussian distribution characterized by the equal-time
covariance matrix $\boldsymbol{S}$, whose elements read
\begin{equation}
S_{ij} = \lim_{t\rightarrow \infty}\langle X_i (t) X_j (t) \rangle_{\eta}. 
\end{equation}  
In addition, one can show that, in the stationary regime, the matrices $\boldsymbol{S}$,  $\boldsymbol{A}$ and $\boldsymbol{D}$ fulfill the relation \cite{godreche2018characterising}
\begin{equation}
  \boldsymbol{AS} + \boldsymbol{SA}^T = 2\boldsymbol{D},
  \label{gfks}
\end{equation}
known as the Sylvester-Lyapunov equation.
One of the aims of our work is 
to introduce a random-matrix ensemble for the covariance matrix $\boldsymbol{S}$ which incorporates the above
constraint. 

Here we focus on MVOU processes that satisfy 
the Onsager reversibility conditions \cite{godreche2018characterising}, expressed in matrix form as follows
\begin{equation}
  \boldsymbol{AD} = \boldsymbol{DA}^T.
  \label{jdu}
\end{equation}
The  reversibility constraint imposes a strong interdependence between the elements of the coupling and diffusion matrices.
For systems that fulfill Eq. (\ref{jdu}), the Sylvester-Lyapunov equation admits the general solution
\begin{equation}
  \boldsymbol{S} = \boldsymbol{A}^{-1}\boldsymbol{D}.
  \label{gfao}
\end{equation}
Thus, in order to put forward a random-matrix ensemble for  $\boldsymbol{S}$, we just need to sample
$\boldsymbol{A}$ and $\boldsymbol{D}$ subject to Eq. (\ref{jdu}), since the constraint imposed by the Sylvester-Lyapunov equation
is automatically fulfilled by Eq. (\ref{gfao}).
We point out that $\boldsymbol{S}$ and $\boldsymbol{D}$ are symmetric matrices, while $\boldsymbol{A}$ is not necessarily symmetric.

In reversible systems, we can also extract information about the lagged covariance in the stationary regime.
Let $R_{ij}(t,s)$ be the two-time response function
\begin{equation}
R_{ij}(t,s) = \frac{\delta \langle  X_i(t) \rangle_{\eta}}{\delta h_j(s)},
\end{equation}
where $h_j(s)$ is a time-dependent external field that linearly couples to $X_j(s)$ in Eq. (\ref{MVOU}).
In the stationary regime, both $R_{ij}(t,s)$ and $S_{ij}(t,s)$
are invariant under time-translation, i.e., $R_{ij}(t,s) = R_{ij}(\tau)$ and $S_{ij}(t,s) = S_{ij}(\tau)$, with $\tau = t-s \geq 0$.
The quantity $S_{ij}(\tau)$ is the lagged covariance between $X_i(t)$ and $X_j(t+\tau)$ in the stationary regime.
In terms of the $N \times N$ matrices $\boldsymbol{R}(\tau)$ and $\boldsymbol{S}(\tau)$, one can show that \cite{godreche2018characterising}
\begin{equation}
\boldsymbol{R}(\tau) = \exp{(-\tau \boldsymbol{A})},
\end{equation}
and
\begin{equation}
  \boldsymbol{S}(\tau) = \exp{(-\tau \boldsymbol{A})} \boldsymbol{S}.
  \label{grws}
\end{equation}
Although Eq. (\ref{grws}) provides an interesting expression
for the lagged covariance $\boldsymbol{S}(\tau)$, the matrices $\boldsymbol{S}$ and $\boldsymbol{A}$
do not share the same eigenvectors. Thus, even if we know the eigenvalues of $\boldsymbol{S}$ and $\boldsymbol{A}$, Eq. (\ref{grws}) does not give access
to the spectrum of $\boldsymbol{S}(\tau)$.

Since MVOU processes are described by linear equations, one can make a change of dynamical variables and derive a more
useful expression for the lagged covariance matrix in the transformed system. Let $\boldsymbol{X}^{\prime} = \boldsymbol{B}^{-1} \boldsymbol{X}$ be
the new vector of dynamical variables, with $\boldsymbol{B}$ an arbitrary matrix. Multiplying  Eq. (\ref{MVOU}) on the left by $\boldsymbol{B}^{-1}$ and using
the decomposition $\boldsymbol{D} = \boldsymbol{B} \boldsymbol{B}^{T}$ of positive-definite matrices, one can show that $\boldsymbol{X}^{\prime}$ fulfills 
\begin{equation}
\label{MVOU11}
d\boldsymbol{X}^{\prime} = -\boldsymbol{A}^{\prime}\boldsymbol{X}^{\prime}dt + \boldsymbol{\eta}^{\prime}\sqrt{dt}, 
\end{equation}
where the new coupling matrix $\boldsymbol{A}^{\prime}$ and the covariance matrix $\boldsymbol{D}^{\prime}$ associated to $\boldsymbol{\eta}^{\prime}$ are, respectively, given
by $\boldsymbol{A}^{\prime} = \boldsymbol{B}^{-1} \boldsymbol{A} \boldsymbol{B}$ and $\boldsymbol{D}^{\prime} = \boldsymbol{I}$, with $\boldsymbol{I}$ denoting the identity matrix.
From the reversibility condition, Eq. (\ref{jdu}), we conclude that $\boldsymbol{A}^{\prime}$ is symmetric. In addition, by inverting
the relation between $\boldsymbol{A}^{\prime}$ and $\boldsymbol{A}$, it is straightforward to verify that
\begin{equation}
  \boldsymbol{A} = \boldsymbol{D} \boldsymbol{J},
  \label{gefs}
\end{equation}  
where $\boldsymbol{J} = (\boldsymbol{B}^T )^{-1} \boldsymbol{A}^{\prime} \boldsymbol{B}^{-1}$ is a symmetric matrix. Hence, the
coupling matrix $\boldsymbol{A}$ of reversible MVOU processes can be always decomposed
as a product between the noise covariance $\boldsymbol{D}$ and a symmetric matrix $\boldsymbol{J}$.
The reversibility condition is automatically fulfilled by the matrix decomposition of Eq. (\ref{gefs}). 

In the transformed system, since the stationary state is characterized by the covariance $\boldsymbol{S}^{\prime} = \boldsymbol{A}^{\prime -1}$, both
the response matrix $\boldsymbol{R}^{\prime}(\tau)$ and the lagged covariance matrix  $\boldsymbol{S}^{\prime} (\tau)$ read
\begin{align}
\boldsymbol{R}^{\prime}(\tau) &= \exp(-\tau \boldsymbol{A}^{\prime}), \nonumber \\
\boldsymbol{S}^{\prime} (\tau) &= \exp(-\tau \boldsymbol{A}^{\prime})\boldsymbol{A}^{ \prime -1}.
\label{hywq}
\end{align}
Therefore, the equal-time covariance $\boldsymbol{S}^{\prime}$, the lagged covariance $\boldsymbol{S}^{\prime} (\tau)$ and
the response $\boldsymbol{R}^{\prime}(\tau)$ share the eigenvectors of the transformed coupling matrix $\boldsymbol{A}^{\prime}$.
Once we know the eigenvalues of $\boldsymbol{A}^{\prime}$, it is straightforward to determine, for instance, the eigenvalues
of the lagged covariance $\boldsymbol{S}^{\prime} (\tau)$ as a function of time $\tau$. 

%%%%%%%%%%%%%%%%%%%%%%%%%%%%%%%%%%%%%%%%%%%%%%%%%%%%%%%%%%%%%%%%%%%%%%%%%%%%%%%%%%
%%%%%%%%%%%%%%%%%%%%%%%%%%%%%%%%%%%%%%%%%%%%%%%%%%%%%%%%%%%%%%%%%%%%%%%%%%%%%%%%%%%%%%

\section{The random-matrix ensemble}
\label{RMT}

We are interested in the spectral properties of the covariance $\boldsymbol{S}$ and
the lagged covariance  $\boldsymbol{S}^{\prime}(\tau)$ that characterize reversible MVOU processes. We will
follow the random-matrix prescription and assume that the coupling
strengths are drawn from an ensemble of random matrices subject to the constraints dictated
by Eqs. (\ref{jdu}) and (\ref{gfao}).

In the previous section, we have shown that $\boldsymbol{A}$ can be
decomposed as $\boldsymbol{A} = \boldsymbol{D} \boldsymbol{J}$, where $\boldsymbol{J}$ is a symmetric matrix
that, in principle, might depend on $\boldsymbol{D}$. Here, we assume that $\boldsymbol{J}$
has the following form
\begin{equation}
\label{adjacencyYY}
\boldsymbol{J} = \mu \boldsymbol{D}^{-1} + g(\boldsymbol{D}) \boldsymbol{K} g(\boldsymbol{D}),
\end{equation}
in which $\mu >0$, $\boldsymbol{K}$ is a symmetric matrix with real-valued entries, and $g(\boldsymbol{D})$ is a matrix function of
the diagonal covariance $\boldsymbol{D}$, with elements
\begin{equation}
D_{ij} = T_i \delta_{ij},
\end{equation}  
where $T_i$ is the local temperature of the dynamical variable $X_i(t)$. Thus, the interaction matrix $\boldsymbol{A}$
and its tranformed version $\boldsymbol{A}^{\prime}$ are given by
\begin{equation}
\label{adjacency}
\boldsymbol{A} = \mu  + \boldsymbol{D}g(\boldsymbol{D}) \boldsymbol{K} g(\boldsymbol{D})
\end{equation}
and
\begin{equation}
\label{adjacency1}
\boldsymbol{A}^{\prime} = \mu  + \boldsymbol{D}^{1/2} g(\boldsymbol{D}) \boldsymbol{K} g(\boldsymbol{D})\boldsymbol{D}^{1/2}.
\end{equation}
The above equations determine the equal-time covariance and the lagged covariance
via Eqs. (\ref{gfao}) and (\ref{hywq}). By tuning $\mu > 0$, we ensure that MVOU processes governed
by Eqs. (\ref{MVOU}) and (\ref{MVOU11}) converge to stationary Gaussian distributions.

We are now ready to formally define the reversible random-matrix ensemble that we study in the following sections.
The diagonal elements of $\boldsymbol{K}$ are zero, while the off-diagonal elements $K_{ij} = K_{ji}$ ($i \neq j$) are independent and identically distributed random variables
drawn from a Gaussian distribution with mean zero and variance $1/N$. In addition, the local
temperatures $T_1,\dots,T_N$ are independent and identically distributed variables sampled from $p(T)$.
The coupling matrix $\boldsymbol{A}$ can be seen as
a deformed Wigner matrix, in which a Gaussian random matrix is multiplied  on both sides
by a generic function of the diagonal
diffusion matrix $\boldsymbol{D}$ and then added to a diagonal matrix.

Our primary aim is to compute the spectral density of $\boldsymbol{S}$. Substituting Eq. (\ref{adjacency}) in Eq. (\ref{gfao}), we obtain
the covariance
\begin{equation}
\boldsymbol{S} = (\mu  + \boldsymbol{D}g(\boldsymbol{D})\boldsymbol{K}g(\boldsymbol{D}))^{-1}\boldsymbol{D},
\end{equation}
and the so-called precision matrix
\begin{equation}
\label{precision}
\boldsymbol{S}^{-1} = \mu \boldsymbol{D}^{-1} + g(\boldsymbol{D})\boldsymbol{K}g(\boldsymbol{D}).
\end{equation}
The spectral density of $\boldsymbol{S}$ is obtained from the spectral density of $\boldsymbol{S}^{-1}$ by a simple
change of variables.

%%%%%%%%%%%%%%%%%%%%%%%%%%%%%%%%%%%%%%%%%%%%%%%%%%%%%%%%%%%%%%%%%%%%%%%%%%%%%%%%%%%%%%
%%%%%%%%%%%%%%%%%%%%%%%%%%%%%%%%%%%%%%%%%%%%%%%%%%%%%%%%%%%%%%%%%%%%%%%%%%%%%%%%%%%%%%%%

\section{Spectral density of the covariances matrices}
\label{specdensity}

In this section, we present analytic results for the spectral densities
of both $\boldsymbol{S}$ and $\boldsymbol{S}^{\prime}(\tau)$ in the limit $N \rightarrow \infty$.
Before discussing our main results, we introduce some useful notation.
Let $\{ \lambda_i (\boldsymbol{X}) \}_{i=1,\dots,N}$ denote the (real) eigenvalues of an $N \times N$ symmetric random matrix $\boldsymbol{X}$. The empirical spectral
density of $\boldsymbol{X}$ is defined as
\begin{equation}
\rho_{X}(\lambda) = \lim_{N \rightarrow \infty} \frac{1}{N} \sum_{j=1}^N \left\langle \delta \left( \lambda - \lambda_j (\boldsymbol{X}) \right) \right\rangle,
\end{equation}  
with $\langle \dots \rangle$ representing the average over the random-matrix ensemble. Introducing the $N \times N$ resolvent matrix
\begin{equation}
  \boldsymbol{G}_{X}(z) = (z \boldsymbol{I} - \boldsymbol{X})^{-1}
  \label{hgdfk}
\end{equation}  
associated to $\boldsymbol{X}$, we obtain $\rho_{X}(\lambda)$ from the resolvent as follows \cite{Metz2019}
\begin{equation}
\rho_{X}(\lambda) = \lim_{\epsilon \rightarrow 0^{+}} \lim_{N \rightarrow \infty} \frac{1}{\pi N} {\rm Im} \langle {\rm Tr} \, \boldsymbol{G}_{X}(z)  \rangle,
\end{equation}  
where $z=\lambda - i \epsilon$.

In the present work, the randomness stems from both heterogeneous temperatures $T_1,\dots,T_N$ and
random couplings $K_{ij}$  (see section \ref{RMT}).
In appendix \ref{replicas}, we explain how to calculate the ensemble average $\langle {\rm Tr} \, \boldsymbol{G}_{S^{-1}}(z)  \rangle$ for
the precision matrix $\boldsymbol{S}^{-1}$
using the replica method \cite{NishimoriBook}.
The spectral density of $\boldsymbol{S}^{-1}$ is given by
\begin{equation}
\label{rho}
    \rho_{S^{-1}}(\lambda) = \frac{1}{\pi} \lim_{\epsilon \to 0^{+}} \text{Im} \left[ \int\limits_{0}^{\infty} dT \frac{p(T)} {\left( z - \frac{\mu}{T} - g^{2}(T) q \right)} \right],
\end{equation}
where the replica-symmetric order-parameter $q$ fulfills the self-consistent equation
\begin{equation}
  \label{order_param}
 q =  \int\limits_{0}^{\infty} dT \frac{p(T) g^{2}(T)}{\left( z - \frac{\mu}{T} - g^{2}(T)q \right)} . 
\end{equation}
Therefore, once we specify the function $g(T)$ and the distribution $p(T)$ of heterogeneous temperatures, we can solve
the fixed-point Eq. (\ref{order_param}) and determine $\rho_{S^{-1}}(\lambda)$.
Given that $\lambda_i (\boldsymbol{S}) = 1/ \lambda_i (\boldsymbol{S}^{-1})$ ($i=1,\dots,N$), the spectral density $\rho_{S}(\lambda)$ of the stationary covariance
matrix $\boldsymbol{S}$ follows from 
\begin{equation}
  \rho_S(\lambda) = \frac{1}{\lambda^2} \rho_{S^{-1}}(1/\lambda).
  \label{cov}
\end{equation}  
Clearly, to determine $\rho_S(\lambda)$, we must solve Eq. (\ref{order_param}) at $z=1/\lambda - i \epsilon$.

Comparing Eqs. (\ref{adjacency1}) and (\ref{precision}), we note that $\boldsymbol{A}^{\prime}$ and $\boldsymbol{S}^{-1}$ have a
similar form. Hence, the replica method in appendix \ref{replicas} can be applied in an analogous way to determine  
the spectral density $\rho_{A^{\prime}}(\lambda)$ of the
transformed interaction matrix $\boldsymbol{A}^{\prime}$.
The final outcome for $\rho_{A^{\prime}}(\lambda)$ reads
\begin{equation}
  \rho_{A^{\prime}}(\lambda) = \frac{1}{\pi} \lim_{\epsilon \to 0^{+}} \text{Im} \left[ \int\limits_{0}^{\infty} dT \frac{p(T)} {\left( z - \mu - T g^{2}(T) q \right)} \right],
  \label{tyty5}
\end{equation}  
where $q$ solves the equation
\begin{equation}
  \label{order_param1}
 q =   \int\limits_{0}^{\infty} dT \frac{p(T) T g^{2}(T)}{\left( z - \mu - T g^{2}(T)q \right)} . 
\end{equation}
According to Eq. (\ref{hywq}), the function $\rho_{A^{\prime}}(\lambda)$ determines the spectral density $\rho_{S^{\prime}(\tau)}(\lambda)$ of the
lagged covariance matrix $\boldsymbol{S}^{\prime}(\tau)$. Indeed, since
$\lambda_j(\boldsymbol{S}^{\prime}(\tau)) = e^{- \tau \lambda_j(\boldsymbol{A}^{\prime}) }/\lambda_j(\boldsymbol{A}^{\prime})$ ($j=1,\dots,N$), we find
the relation
\begin{equation}
  \rho_{S^{\prime}(\tau)}(\lambda) = \frac{\lambda^{\prime}}{\lambda \left( 1 + \tau \lambda^{\prime} \right)} \rho_{A^{\prime}}(\lambda^{\prime}),
  \label{cov1}
\end{equation}  
where $\lambda^{\prime} =\lambda^{\prime}(\lambda) $ is determined from the solutions of the fixed-point equation
\begin{equation}
  e^{- \tau \lambda^{\prime}} = \lambda \lambda^{\prime}.
  \label{ww2}
\end{equation}  

%%%%%%%%%%%%%%%%%%%%%%%%%%%%%%%%%%%%%%%%%%%%%%%%%%%%%%%%%%%%%%%%%%%%%%%%%%%%%%%%%%%%
%%%%%%%%%%%%%%%%%%%%%%%%%%%%%%%%%%%%%%%%%%%%%%%%%%%%%%%%%%%%%%%%%%%%%%%%%%%%%%%%%%%%

\section{Results}
\label{secresults}

In this section, we discuss our results for the spectral density of both lagged and stationary covariance matrices obtained
from Eqs. (\ref{cov}) and (\ref{cov1}).
In order to solve the equations for the spectral densities, we need to specify the function $g(T)$ and the
distribution $p(T)$ of temperatures. Here, we choose $g(T)=T^{-\alpha}$, with $\alpha \in [0,1]$, and the entries of the coupling matrices $\boldsymbol{A}$ and $\boldsymbol{A}^{\prime}$ assume the form
\begin{align}
  A_{ij} &= \mu \delta_{ij} +  T_i^{1-\alpha} K_{ij} T_j^{-\alpha}  , \label{uyq} \\
 A_{ij}^{\prime} &= \mu \delta_{ij} + T_i^{\frac{1}{2}-\alpha} K_{ij}  T_j^{\frac{1}{2} -\alpha}. \label{uyq1} 
\end{align}
The  matrix $\boldsymbol{A}$ is asymmetric, with $A_{ij}$ representing the interaction strength or
influence of $X_j(t)$ on $X_i(t)$.
The exponent $\alpha$ shapes the role of the local
temperatures on the pairwise interactions. For instance, when $\alpha=1$,  $A_{ij}$ is weighted
by $1/T_j$, which means that a variable $X_j(t)$ with a high temperature $T_j$ will have a weak influence
on the rest of the system.

We are also interested in the stability of MVOU processes.
The dynamics of Eq. (\ref{MVOU}) evolves to stable stationary states if all eigenvalues of the
precision matrix $\boldsymbol{S}^{-1}$ are non-negative \cite{godreche2018characterising,fyodorov2023fluctuation}. If we order these
eigenvalues  as $\lambda_1(\boldsymbol{S}^{-1}) < \lambda_2(\boldsymbol{S}^{-1}) < \dots < \lambda_N(\boldsymbol{S}^{-1})$, MVOU processes
are stable provided $\lambda_1(\boldsymbol{S}^{-1}) \geq 0$. Therefore, the lower spectral edge of $\rho_{S^{-1}}(\lambda)$ determines
whether the system is stable  in the limit $N \rightarrow \infty$. At marginal stability, the lower spectral
edge of $\rho_{S^{-1}}(\lambda)$ touches zero, and the decay of $\rho_S(\lambda)$ for large $\lambda$ is obtained from
the functional behavior of $\rho_{S^{-1}}(\lambda)$ as $\lambda \rightarrow 0^{+}$ (see Eq. (\ref{cov})).

We can also  characterize the stability of MVOU processes in terms
of the macroscopic parameter
\begin{equation}
  m(t) = \lim_{N \rightarrow \infty} \frac{1}{N} \sum_{i=1}^N \left\langle X_i(t) \right\rangle_{\eta} .
  \label{mt}
\end{equation}  
For a stable MVOU process ($\lambda_1(\boldsymbol{S}^{-1}) \geq  0$), $m(t)$ relaxes to the
trivial fixed-point $m=0$. Conversely, for
an unstable MVOU process, $m(t)$ diverges.
Below, we present results for an homogeneous temperature and two cases of heterogeneous
temperatures: a bimodal and a uniform distribution $p(T)$.

%%%%%%%%%%%%%%%%%%%%%%%%%%%%%%%%%%%%%%%%%%%%%%%%%%%%%%%%%%%%%%%%%%%%%%%%%%%%%%%%%%%%%%%%
%%%%%%%%%%%%%%%%%%%%%%%%%%%%%%%%%%%%%%%%%%%%%%%%%%%%%%%%%%%%%%%%%%%%%%%%%%%%%%%%%%%%%%%%%%
  
\subsection{Homogeneous temperatures}
  
Let us consider the analytically solvable
case of a homogeneous temperature, where $T_i = T \,\,\, \forall \, i$. Substituting
$p(T^{\prime})=\delta(T^{\prime}-T)$ in Eq. (\ref{order_param}) and solving the resulting quadratic equation for $q$, we obtain
\begin{equation}
q = \frac{1}{2 g^{2}(T)} \left(z - \frac{\mu}{T} + \sqrt{\left( \frac{\mu}{T} - z  \right)^2 - \Lambda^2  }   \right),
\end{equation}  
with $\Lambda = 2 g^{2}(T)$.
Inserting this result into Eq. (\ref{rho}) and taking the limit $\epsilon \rightarrow 0^{+}$, we derive the Wigner semicircle law for the spectral
density of the precision matrix,
\begin{equation}
  \rho_{S^{-1}}(\lambda) =
\begin{dcases}
    \frac{2}{\pi \Lambda^2} \sqrt{\Lambda^2 - \left( \frac{\mu}{T}  - \lambda \right)^2 } & \text{if } \,\, \lambda \in \left(\lambda_{-},\lambda_{+} \right), \\
    0              & \text{otherwise},
    \label{djfg}
\end{dcases}
\end{equation}  
where the spectral edges read $\lambda_{\pm} = \frac{\mu}{T} \, \pm \, \Lambda$. For stable MVOU processes, where $\lambda_{-} > 0$, the spectral density
of $\boldsymbol{S}$ is given by
\begin{equation}
\rho_{S}(\lambda) =
\begin{dcases}
       \frac{2}{ \pi \Lambda^2 \lambda^2 } \sqrt{\Lambda^2 - \left( \frac{\mu}{T}  - \frac{1}{\lambda} \right)^2 }  & \!\!\!\!\! \text{if } \,\, \lambda \in \left(\lambda_{+}^{-1},\lambda_{-}^{-1} \right), \\
    0              & \!\!\!\!\! \text{otherwise}.
\end{dcases}
\label{hyte}
\end{equation}  
The above expression follows from Eqs. (\ref{cov}) and (\ref{djfg}).

\begin{figure}[h!]
    \centering
    \includegraphics[scale=0.52]{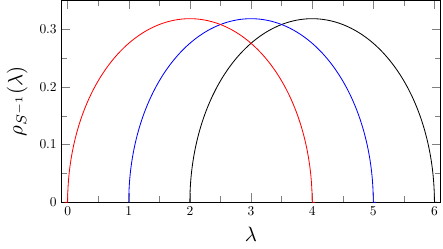}
    \includegraphics[scale=0.52]{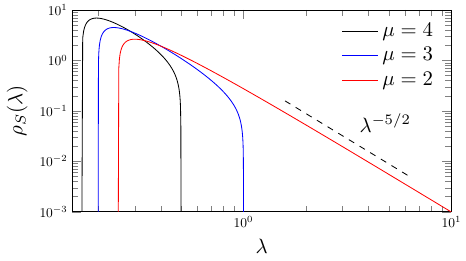}
    \caption{Spectral density of the precision matrix $\boldsymbol{S}^{-1}$ (Eq. (\ref{djfg})) and of the covariance matrix $\boldsymbol{S}$ for stationary MVOU processes
      interacting through Eq. (\ref{uyq}) with $\alpha=1$. The temperatures are equal to $T_i=1 \,\, \forall i$. In the stable
      regime ($\mu > 2$), the spectral density $\rho_{S}(\lambda)$ is given by Eq. (\ref{hyte}), while it
      exhibits a power-law tail at marginal stability ($\mu=2$).}
    \label{hefj6}
\end{figure}

We note that $\rho_{S}(\lambda)$ has a bounded support if the
MVOU process is stable ($\lambda_{-} > 0$).
However, as the system approaches marginal stability ($\lambda_{-} \rightarrow 0^{+}$), the support  becomes unbounded
and the distribution $\rho_{S}(\lambda)$ develops a power-law tail $\rho_{S}(\lambda) \propto \lambda^{-5/2}$ for large $\lambda$. This means
that the variables $X_1(t),\dots,X_N(t)$ are strongly correlated at marginal stability. In Fig. \ref{hefj6}, we illustrate the behavior
of $\rho_{S^{-1}}(\lambda)$ and $\rho_{S}(\lambda)$ across the stability transition.
By setting $\lambda_{-} = 0$, we find the critical value $\mu_c = T \Lambda$ at which the system is marginally stable.
We also point out that the shape of $\rho_{S}(\lambda)$ is similar to the one in \cite{Laloux1999}, with the difference
that our results follow from
a random-matrix model for the interaction matrix $\boldsymbol{A}$, while in \cite{Laloux1999} the spectral density of $\boldsymbol{S}$
is derived from the Wishart random-matrix ensemble.

From Eqs. (\ref{tyty5}) and (\ref{order_param1}), one can show that $\rho_{A^{\prime}}(\lambda)$ also follows the Wigner
semicircle law, which yields
the spectral density of the lagged covariance matrix $S^{\prime}(\tau)$ of the stable MVOU process described by Eq. (\ref{MVOU11}),
\begin{equation}
\rho_{S^{\prime}(\tau)}(\lambda) =
\begin{dcases}
  F(\lambda,\lambda^{\prime})
  \sqrt{T^2 \Lambda^2 - \left( \mu  - \lambda^{\prime} \right)^2 }  & \!\!\!\!\! \text{if }  \lambda \in \left(\gamma_{+},\gamma_{-} \right) \\
    0              & \!\!\!\!\! \text{otherwise}.
\end{dcases}
\label{ytda}
\end{equation}  
The function $F(\lambda,\lambda^{\prime})$ is defined as
\begin{equation}
F(\lambda,\lambda^{\prime}) = \frac{2 \lambda^{\prime}}{ \pi \Lambda^2 T^2  \lambda (1 + \tau \lambda^{\prime})},
\end{equation} 
while the variable $\lambda^{\prime}$ solves Eq. (\ref{ww2}). The upper and lower spectral edges of $\rho_{S^{\prime}(\tau)}(\lambda)$ are given by
\begin{equation}
\gamma_{\pm} = \frac{\exp{\left[ -\tau \left( \mu \pm  T \Lambda   \right) \right]} }{  \mu \pm T \Lambda }.
\end{equation}  
The finite and positive support of $\rho_{S^{\prime}(\tau)}(\lambda)$ reflects the stability of the MVOU process.

\begin{figure}[H]
    \centering
    \includegraphics[scale=0.51]{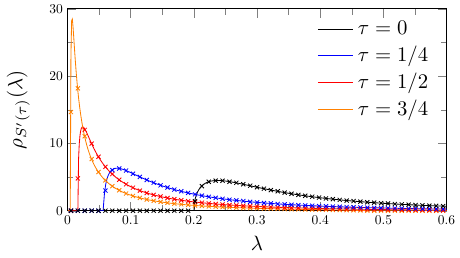}
    \includegraphics[scale=0.51]{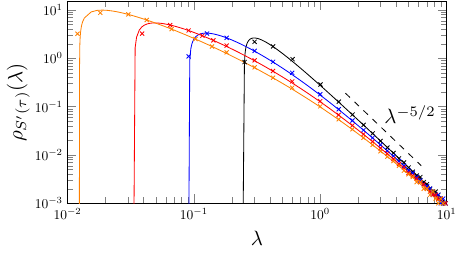}
    \caption{Spectral density $\rho_{S^{\prime}(\tau)}(\lambda)$ of the lagged covariance matrix $\boldsymbol{S}^{\prime}(\tau)$ for
      stationary MVOU processes described by Eq. (\ref{MVOU11}).  The coupling matrix
      is given by Eq. (\ref{uyq1}), with $\alpha=1$ and a homogeneous temperature $T_i=1 \,\, \forall i$. The left
      panel shows $\rho_{S^{\prime}(\tau)}(\lambda)$ in the stable regime ($\mu=3$), while the right panel illustrates
      the power-law decay of $\rho_{S^{\prime}(\tau)}(\lambda)$ at marginal stability ($\mu=2$). The solid lines are obtained
      from Eq. (\ref{ytda}), while the symbols are numerical diagonalization results derived from an ensemble
      of $10$ matrices $\boldsymbol{S}^{\prime}(\tau)$ with $N= 10^{4}$.}
    \label{hej7}
\end{figure}
As $\mu  \rightarrow \mu_{c}^{+}$, the upper spectral edge $\gamma_{-}$ diverges, and $\rho_{S^{\prime}(\tau)}(\lambda)$ decays
as $\rho_{S^{\prime}(\tau)}(\lambda) \propto \lambda^{-5/2}$ for large $\lambda$. Figure \ref{hej7} illustrates the effect of the time-difference $\tau$ on the
spectral density $\rho_{S^{\prime}(\tau)}(\lambda)$ for both  stable ($\mu > \mu_c$) and marginally stable ($\mu=\mu_c$) MVOU
processes. As $\tau$ increases, $\rho_{S^{\prime}(\tau)}(\lambda)$ develops a peak close to $\lambda=0$, corresponding to uncorrelated
dynamical variables in the stationary state. In the marginally stable regime, $\rho_{S^{\prime}(\tau)}(\lambda)$ also exhibits an excess of modes
around zero, but large correlations among the dynamical variables lead to the power-law decay in  $\rho_{S^{\prime}(\tau)}(\lambda)$.
For $\tau \rightarrow 0$, $\rho_{S^{\prime}(\tau)}(\lambda)$ converges to the spectral density of the
equal-time covariance $\boldsymbol{S}^{\prime}$, which characterizes  the stationary states of Eq. (\ref{MVOU11}).

%%%%%%%%%%%%%%%%%%%%%%%%%%%%%%%%%%%%%%%%%%%%%%%%%%%%%%%%%%%%%%%%%%%%%%%%%%%%%%%%%%%%%%%%
%%%%%%%%%%%%%%%%%%%%%%%%%%%%%%%%%%%%%%%%%%%%%%%%%%%%%%%%%%%%%%%%%%%%%%%%%%%%%%%%%%%%%%%

\subsection{Bimodal temperature distribution}
In this subsection, we present results for the bimodal distribution of temperatures
\begin{equation}
\label{dist_bimod}
    p_{\rm b}(T) = p \delta(T-T_0) + (1-p) \delta(T-T_0 - \delta),
\end{equation}
with $T_0 > 0$ and $\delta > 0$. The parameter $p \in [0,1]$ determines the fraction of dynamical
variables $\{ X_i(t) \}_{i=1,\dots,N}$ in the lowest temperature $T_0$.
Below, we exploit how bimodal temperatures impact the stability transition of MVOU processes and the spectral densities of the covariance matrices.

\begin{figure}[h!]
    \centering
    \includegraphics[scale=0.8]{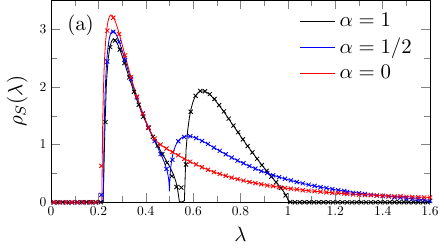}
    \includegraphics[scale=0.8]{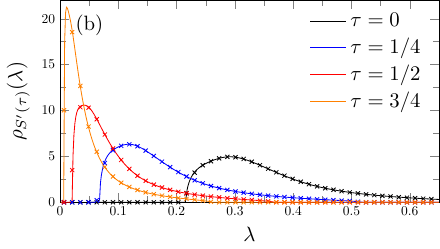}
\caption{(a) Spectral density $\rho_{S}(\lambda)$ of the covariance matrix $\boldsymbol{S}$ for stationary
  MVOU processes interacting through Eq. (\ref{uyq}) with $\mu=3$ and different values of $\alpha$.
 (b) Spectral density $\rho_{S^{\prime}(\tau)}(\lambda)$
  of the lagged covariance matrix $S^{\prime}(\tau)$ for stationary MVOU processes (see Eq. (\ref{MVOU11})) interacting
  through Eq. (\ref{uyq1}) with $\alpha=1$ and $\mu=3$.
  The temperatures in both panels follow a bimodal distribution with $T_0=\delta=1$ and $p=1/2$ (see Eq. (\ref{dist_bimod})).
  The solid lines are obtained from the solutions of Eqs. (\ref{order_param}) and (\ref{order_param1}) with $\epsilon=10^{-3}$, while
  the symbols are numerical diagonalization results derived from an ensemble of $10$ covariance
  matrices with $N = 10^4$.}
    \label{ESD_bimod}
\end{figure}

For heterogeneous temperatures, where the variance
of $p(T)$ is finite, we obtain the spectral densities of $\boldsymbol{S}$ and $\boldsymbol{S}^{\prime}(\tau)$
by numerically solving the self-consistent Eqs.  (\ref{order_param}) and (\ref{order_param1}), respectively.
In Fig. \ref{ESD_bimod}, we compare numerical diagonalization results with our theoretical findings
for $\rho_{S}(\lambda)$ and $\rho_{S^{\prime}(\tau)}(\lambda)$ in the limit $N \rightarrow \infty$.
The excellent agreement between the two approaches validates our theoretical predictions. The results in Fig. \ref{ESD_bimod} are for
stable MVOU processes, where $\rho_{S}(\lambda)$ is supported on a finite, positive interval.

\begin{figure}[h!]
    \centering
    \includegraphics[scale=0.8]{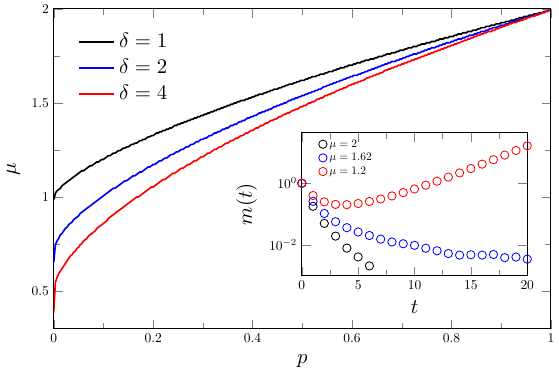}
    \includegraphics[scale=0.51]{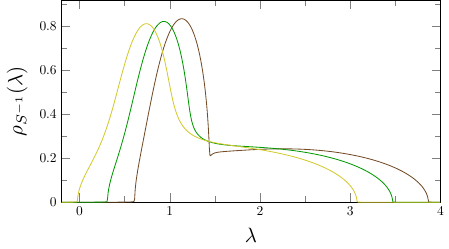}
    \includegraphics[scale=0.51]{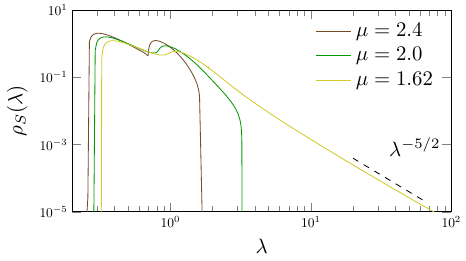}   
    \caption{Main panel: stability diagram $(\mu,p)$ of stationary MVOU processes with $\alpha=1$ (see Eq. (\ref{uyq})). 
      The temperatures are drawn from the bimodal distribution of Eq. (\ref{dist_bimod}) with $T_0=1$. The MVOU process is stable above the solid lines, unstable
      below them, and marginally stable at the lines.
      The inset displays the dynamics of $m(t)$, Eq. (\ref{mt}), for $p=1/2$, $\delta=1$, and different $\mu$. These results follow from the numerical integration
      of Eq. (\ref{MVOU}) for an ensemble of $N=10^3$ dynamical variables. Lower panels: spectral
      densities of the covariance and precision matrices, $\boldsymbol{S}$ and $\boldsymbol{S}^{-1}$, across
      the stability transition for $T_0=\delta=1$ and $p=1/2$. These results are obtained from the solutions
      of Eq. (\ref{order_param}) with $\epsilon=10^{-6}$.
    } 
    \label{PD_bimod}
\end{figure}

As $\mu$ decreases, MVOU processes interacting through Eq. (\ref{uyq}) become unstable. Figure \ref{PD_bimod} shows the stability
diagram $(\mu,p)$ for $\alpha=1$ and a bimodal temperature distribution. Since  for $\alpha=1$ the couplings $A_{ij}$ are weighted
by $1/T_j$, the system becomes more stable as $p$ decreases, due to the larger number of weaker interactions
among the dynamical variables. For the same reason, a larger temperature difference $\delta$ also promotes stability.
The system is marginally stable at the solid lines, which are determined by the values of $(\mu,p)$ where 
the lower spectral edge of $\rho_{S^{-1}}(\lambda)$ is zero.
The inset in Fig. \ref{PD_bimod} shows numerical simulations confirming that the macroscopic variable $m(t)$
relaxes to zero when the system is stable, while it diverges in the unstable regime.

The lower panels in Fig. \ref{PD_bimod} show the behavior of the  spectral densities $\rho_{S}(\lambda)$ and
$\rho_{S^{-1}}(\lambda)$ across the stability transition. In the stable regime, $\rho_{S}(\lambda)$ is supported on a finite
interval and it exhibits two maxima, reflecting the bimodal shape of  $p_{\rm b}(T)$.
At marginal stability, we have numerically confirmed
that $\rho_{S^{-1}}(\lambda) \propto \sqrt{\lambda}$ as $\lambda \rightarrow 0^{+}$, independently of $\alpha$. This behavior
leads to an unbounded density $\rho_{S}(\lambda)$ with a power-law tail $\rho_{S}(\lambda) \propto \lambda^{-5/2}$ for large $\lambda$.

%%%%%%%%%%%%%%%%%%%%%%%%%%%%%%%%%%%%%%%%%%%%%%%%%%%%%%%%%%%%%%%%%%%%%%%%%%%%%%%%%%%%%
%%%%%%%%%%%%%%%%%%%%%%%%%%%%%%%%%%%%%%%%%%%%%%%%%%%%%%%%%%%%%%%%%%%%%%%%%%%%%%%%%%%%%%

\subsection{Uniform temperature distribution}

\begin{figure}[h!]
    \centering
    \includegraphics[scale=0.8]{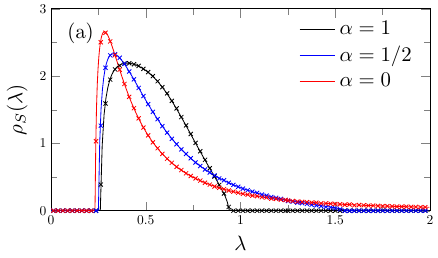}
    \includegraphics[scale=0.8]{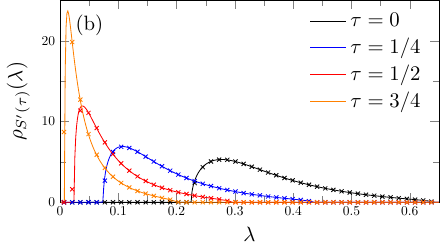}
    \caption{(a) Spectral density $\rho_{S}(\lambda)$ of the covariance matrix $\boldsymbol{S}$ for stationary
      MVOU processes interacting through Eq. (\ref{uyq}) with $\mu=3$ and different $\alpha$. (b) Spectral density $\rho_{S^{\prime}(\tau)}(\lambda)$
      of the lagged covariance matrix $S^{\prime}(\tau)$ for stationary MVOU processes (see Eq. (\ref{MVOU11})) interacting
  through Eq. (\ref{uyq1}) with $\alpha=1$ and $\mu=3$.
   The temperatures in both panels follow an uniform distribution with $T_M=3/2$ and $\Delta=1$ (see Eq. (\ref{uni11})). The solid lines
  are derived from the solutions of Eqs. (\ref{order_param}) and (\ref{order_param1})
      with $\epsilon=10^{-3}$, while the symbols are numerical
      diagonalization results obtained from an ensemble of $10$ covariance matrices with $N = 10^4$.}
    \label{spec_u}
\end{figure}

In this subsection, we derive results for the uniform distribution of temperatures
\begin{equation}
    p_{\rm u}(T)= 
\begin{dcases}
    \Delta^{-1} & \text{if } \,\, T \in \left( T_M - \Delta/2,T_M + \Delta/2 \right), \\
    0              & \text{otherwise},
\end{dcases}
\label{uni11}
\end{equation}
where $T_M > 0$ and $\Delta \in (0,2 \, T_M)$ are, respectively, the center and the width of $p_{\rm u}(T)$.
In Fig. \ref{spec_u}, we compare our analytic results for $\rho_{S}(\lambda)$ and $\rho_{S^{\prime}(\tau)}(\lambda)$ with
numerical diagonalization results of finite covariance matrices for temperatures drawn from Eq. (\ref{uni11}). The results in Fig. \ref{spec_u} once
more confirm the exactness of our theoretical findings for $N \rightarrow \infty$.

\begin{figure}
    \centering
    \includegraphics[scale=0.8]{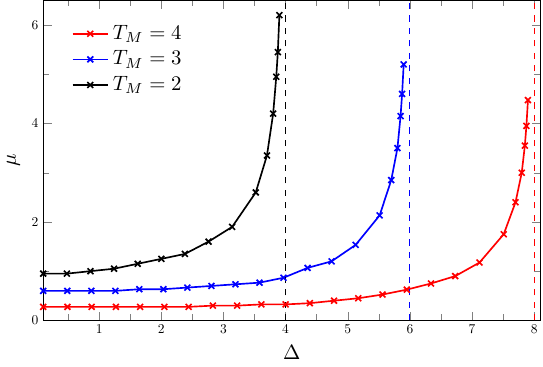}
    \includegraphics[scale=0.51]{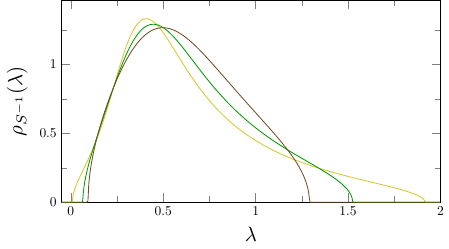}
    \includegraphics[scale=0.51]{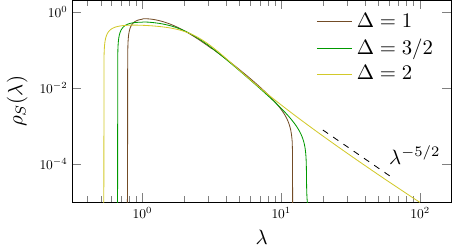}   
    \caption{Main panel: stability diagram $(\mu,\Delta)$ of stationary MVOU processes with $\alpha=1$ (see Eq. (\ref{uyq})). The temperatures
      follow the uniform distribution of Eq. (\ref{uni11}). The symbols identify points of the stability transition, while the solid lines
      are a guide to the eye. The vertical dashed lines mark the maximum value $\Delta = 2 T_M$.
      The MVOU process is stable above the transition points, unstable below them, and marginally
      stable at the points. Lower panels: spectral densities of the covariance and precision matrices, $\boldsymbol{S}$ and $\boldsymbol{S}^{-1}$, for $T_M=2$, $\mu=1.2$, and
      increasing values of $\Delta$. These results are obtained from the solutions
      of Eq. (\ref{order_param}) with $\epsilon=10^{-6}$.
    }
    \label{PD_unif}
\end{figure}

Figure \ref{PD_unif} shows the stability diagram $(\mu,\Delta)$ for stationary MVOU processes interacting
according to Eq. (\ref{uyq}) with $\alpha=1$. The temperatures follow the uniform
distribution specified in Eq. (\ref{uni11}). As the width $\Delta$ increases towards its
maximum $2 \, T_M$, the system becomes less stable, due to
the growing number of strong pairwise couplings arising from small temperatures.
Additionally, Fig. \ref{PD_unif} illustrates the evolution of $\rho_{S}(\lambda)$ and $\rho_{S^{-1}}(\lambda)$
as $\Delta$ increases from the stable regime to the marginal stability line. The effect of increasing temperature fluctuations
is to broaden the support of the spectral densities. At marginal stability, $\rho_{S}(\lambda)$
exhibits once more the power-law tail $\rho_{S}(\lambda) \propto \lambda^{-5/2}$, due to the functional behavior $\rho_{S^{-1}}(\lambda) \propto \sqrt{\lambda}$
close to the lower spectral edge $\lambda=0$.

%%%%%%%%%%%%%%%%%%%%%%%%%%%%%%%%%%%%%%%%%%%%%%%%%%%%%%%%%%%%%%%%%%%%%%%%%%%%%%%%%%%%%%%%
%%%%%%%%%%%%%%%%%%%%%%%%%%%%%%%%%%%%%%%%%%%%%%%%%%%%%%%%%%%%%%%%%%%%%%%%%%%%%%%%%%%%%%%%

\section{Discussion}

We have introduced a random matrix model for covariance matrices of reversible multivariate Ornstein-Uhlenbeck
processes (MVOU), in which the ensemble is constrained by the solutions
of the Sylvester-Lyapunov equation.
In contrast to traditional random-matrix ensembles of the covariance \cite{Laloux1999,burda2004free,Burda2004}, where
the statistics of the matrix elements is independent of the interactions and diffusion terms, here
the covariance naturally follows from the distribution of coupling strengths
and temperatures characterizing MVOU processes, representing an alternative family of null
models for the empirical correlations in stochastic complex systems.
In addition, by making a linear transformation
of the dynamical variables, we have shown how the lagged covariance matrix becomes a simple function of the coupling matrix in the transformed
system. Thus, our ensemble enables to study the effect of random interactions and heterogeneous temperatures on the spectral density of both the equal-time
covariance and the lagged covariance matrices that characterize the stationary states of reversible MVOU processes.

Building on the replica method of spin-glass theory \cite{NishimoriBook}, we have derived an exact equation for the spectral density of the equal-time covariance
matrix of the stationary states.
We have shown that the stationary states
undergo a transition between stability and instability, which manifests itself on the shape of the spectral density. In the stable
regime, all eigenvalues of the covariance are positive and its spectral density is bounded, while negative eigenvalues emerge
in the unstable phase. Interestingly, at marginal stability, the spectral density $\rho_{S}(\lambda)$ becomes unbounded, decaying as
a power-law $\rho_{S}(\lambda) \propto \lambda^{-5/2}$  for large eigenvalues. The exponent of the power-law tail is independent
of the details defining the model, such as the distribution of temperatures, indicating an universal behavior of the covariance
at marginal stability. Based on the spectral properties of the covariance matrix, we have derived phase diagrams that illustrate
how temperature fluctuations influence the stability of the system.

The flexibility of the model introduced here allows to explore how more structured interactions in MVOU processes influence
the spectral density of the covariance matrix.
An important question in this context is whether the general features of the spectral density across the stability transition
remain valid in the more realistic scenario of sparse interactions \cite{Metz2019}. Additionally, to establish the power-law decay as a robust
property of the covariance spectral density, it is crucial to study how the nonlinearity in the dynamics of complex systems \cite{Sompolinsky1988,Bunin2017,Marti2018}
impacts the spectral properties of the covariance at marginal stability.
Finally, we highlight
that our random-matrix ensemble offers an alternative null model for the empirical covariance
of complex systems. We hope this will stimulate comparisons between our theoretical predictions and
empirical data.

\begin{acknowledgments}  
L. S. F. acknowledges a fellowship from CNPq/Brazil. F. L. M. thanks CNPq/Brazil for financial support.
\end{acknowledgments}

\appendix

%%%%%%%%%%%%%%%%%%%%%%%%%%%%%%%%%%%%%%%%%%%%%%%%%%%%%%%%%%%%%%%%%%%%%%%%%%%%%%%
%%%%%%%%%%%%%%%%%%%%%%%%%%%%%%%%%%%%%%%%%%%%%%%%%%%%%%%%%%%%%%%%%%%%%%%%%%%%%%%

\section{Replica method for the spectral density of the precision matrix}
\label{replicas}
In this appendix, we explain how to employ the replica method of disordered systems to obtain an analytic
expression for the spectral density of the precision matrix $\boldsymbol{S}^{-1}$ in the limit $N \rightarrow \infty$. The derivation
of the spectral density of the transformed matrix $\boldsymbol{A}^{\prime}$ is completely analogous, and here we will state only the final result.

Let us introduce the $N \times N$ resolvent matrix of $\boldsymbol{S}^{-1}$,
\begin{equation}
\boldsymbol{G}_{S^{-1}}(z) = (z \boldsymbol{I} - \boldsymbol{S}^{-1})^{-1},
\end{equation}  
where $\boldsymbol{I}$ is the identity matrix and $z = \lambda - i \epsilon$, with the regularizer $\epsilon > 0$. The precision matrix is defined
in Eq. (\ref{precision}). The empirical
spectral density of $\boldsymbol{S}^{-1}$ follows from
\begin{equation}
  \rho_{S^{-1}}(\lambda) = \lim_{\epsilon \rightarrow 0^{+}} \lim_{N \rightarrow \infty} \frac{1}{\pi N} {\rm Im} \left\langle {\rm Tr} \, \boldsymbol{G}_{S^{-1}}(z) \right\rangle ,
  \label{uydqs}
\end{equation}  
with $\langle \dots \rangle$ denoting the ensemble average over the temperatures $T_1,\dots,T_N$ and the off-diagonal entries of $\boldsymbol{K}$. The
latter follow a Gaussian distribution with mean zero and variance $1/N$.

To proceed further, we map the problem of computing the ensemble average $\left\langle {\rm Tr} \, \boldsymbol{G}_{S^{-1}}(z) \right\rangle$ into a statistical
mechanics calculation \cite{Edwards1976}. By using the identity
\begin{equation}
{\rm Tr} \boldsymbol{G}_{S^{-1}}(z) = - 2 \frac{\partial}{\partial z} \ln \left[ \det \left( z \boldsymbol{I} - \boldsymbol{S}^{-1}   \right)  \right]^{-1/2},
\end{equation}  
and representing $\left[ \det \left( z - \boldsymbol{S}^{-1}   \right)  \right]^{-1/2}$ as a Gaussian integral over real
variables $\boldsymbol{\phi} = (\phi_1,\dots\phi_N)^{T}$, we rewrite the above expression as
\begin{equation}
{\rm Tr} \boldsymbol{G}_{S^{-1}}(z) = -  2 \frac{\partial}{\partial z} \ln \mathcal{Z}(z),
\end{equation}  
where
\begin{equation}
  \mathcal{Z}(z) = \int\limits_{-\infty}^{\infty} \Bigg( \prod\limits_{j=1}^{N} d \phi_j   \Bigg) \exp{\left[ - \frac{i}{2} \boldsymbol{\phi}^{T}
      \left( z \boldsymbol{I} - \boldsymbol{S}^{-1}  \right) \boldsymbol{\phi} \right] }
  \label{hft1}
\end{equation}  
is the partition function associated to the random-matrix model. Thus, the problem of computing the ensemble average $\left\langle {\rm Tr} \, \boldsymbol{G}_{S^{-1}}(z) \right\rangle$ boils down
to calculate the averaged free-energy $\langle \ln \mathcal{Z}(z) \rangle$.

To calculate the average over the random-matrix ensemble, we use the replica method \cite{NishimoriBook}, which is based on the following identity
\begin{equation}
  \langle \ln \mathcal{Z}(z) \rangle = \lim_{n \rightarrow 0} \frac{1}{n} \ln \langle \mathcal{Z}^{n}(z) \rangle.
  \label{fdsa1}
\end{equation}  
The strategy is to compute first the average $\langle \mathcal{Z}^{n}(z) \rangle$ for a positive integer $n$, and then reconstruct $\langle \ln \mathcal{Z}(z) \rangle$ by taking
the limit $n \rightarrow 0$, according to Eq. (\ref{fdsa1}). By substituting Eq. (\ref{precision}) in Eq. (\ref{hft1}) and then performing the average
of the replicated partition function $\mathcal{Z}^{n}(z)$ over the Gaussian distributed off-diagonal elements $K_{ij}$, we obtain
\begin{align}
  &\langle \mathcal{Z}^{n}(z) \rangle \simeq \Bigg\langle \Bigg( \prod_{j=1}^N d \boldsymbol{\phi}_j  \Bigg)
  \exp{\left[ - \frac{i}{2} \sum\limits_{i=1}^N \left( z- \frac{\mu}{T_i} \right) \boldsymbol{\phi}_i^2   \right]  }  \nonumber \\
  &\times \exp{\Bigg[- \frac{N}{4} \sum_{\alpha,\beta=1}^n \left(  \frac{1}{N} \sum_{i=1}^N  g^2(T_i) \phi_{i}^{\alpha} \phi_{i}^{\beta}   \right)^2   \Bigg] } \Bigg\rangle_{T_1,\dots,T_N},
  \label{tkdl}
\end{align}  
where $\boldsymbol{\phi}_i = (\phi_{i}^{1},\dots,\phi_{i}^{n})^{T}$ is the $n$-dimensional vector in the replica space, and $\langle \dots \rangle_{T_1,\dots,T_N}$ represents
the average over the temperatures $T_1,\dots,T_N$. We have
neglected terms of $O(N^{0})$ in the exponent of Eq. (\ref{tkdl}), since these are subleading contributions in the limit $N \rightarrow \infty$.
Using the Hubbard-Stratonovich transformation,
\begin{align}
 & \sqrt{\pi N} \exp{\left[ - \frac{N}{4} \left(  \frac{1}{N} \sum\limits_{i=1}^N g^{2}(T_i) \phi_{i}^{\alpha} \phi_{i}^{\beta}     \right)^2 \right] } = \nonumber \\
 & \int d q_{\alpha \beta} \, \exp{\left( - \frac{q_{\alpha \beta}^2}{N} + \frac{i q_{\alpha \beta}}{N} \sum_{i=1}^N g^{2}(T_i) \phi_{i}^{\alpha} \phi_{i}^{\beta}  \right)},
\end{align}  
we introduce the order-parameters $\{ q_{\alpha \beta} \}_{\alpha,\beta=1,\dots,N}$, which enables to decouple the
variables $\boldsymbol{\phi}_1,\dots,\boldsymbol{\phi}_N$  at different sites. Additionally, by rescaling
$q_{\alpha \beta}$ as $q_{\alpha \beta} \rightarrow N q_{\alpha \beta}$, we can rewrite Eq. (\ref{tkdl}) as follows
\begin{equation}
  \langle \mathcal{Z}^{n}(z) \rangle \simeq \int \left( \prod\limits_{\alpha, \beta=1}^n d q_{\alpha \beta}  \right) e^{- N \Phi(\{ q_{\alpha \beta}  \} )},
  \label{hdbsq}
\end{equation}  
where
\begin{equation}
   \Phi(\{ q_{\alpha \beta}   \} ) = \sum_{\alpha,\beta=1}^n q_{\alpha \beta}^{2} - \ln{\left\langle \int d \boldsymbol{\phi} \, e^{ H_T(\boldsymbol{\phi})  }   \right\rangle_T }.
\end{equation}  
The function $H_T(\boldsymbol{\phi})$,
\begin{equation}
H_T(\boldsymbol{\phi}) = - \frac{i}{2} \left( z - \frac{\mu}{T}  \right) \boldsymbol{\phi}^2
+ i g^{2}(T) \sum\limits_{\alpha,\beta=1}^n q_{\alpha \beta} \phi^{\alpha} \phi^{\beta},
\end{equation}  
can be interpreted as an effective single-site Hamiltonian.
When writing Eq. (\ref{hdbsq}), we have ignored constants that do not contribute to $\langle \mathcal{Z}^{n}(z) \rangle$ in the limit $N \rightarrow \infty$.

We can now evaluate the integral in Eq. (\ref{hdbsq}) using the saddle-point method. In the limit $N \rightarrow \infty$, $\langle \mathcal{Z}^{n}(z) \rangle $ is asymptotically
given by
\begin{equation}
  \langle \mathcal{Z}^{n}(z) \rangle \sim e^{- N \Phi(\{ q_{\alpha \beta}  \} )},
  \label{rtrt77}
\end{equation} 
where the order-parameters $\{ q_{\alpha \beta} \}_{\alpha,\beta=1,\dots,N}$ fulfill the saddle-point equations
\begin{equation}
  q_{\alpha \beta} = \frac{i}{2} \frac{ \left\langle  g^{2}(T) \int d \boldsymbol{\phi} \, \phi^{\alpha} \phi^{\beta} e^{H_T (\boldsymbol{\phi}) }   \right\rangle_T   }
  {\left\langle \int d \boldsymbol{\phi} \, e^{  H_T (\boldsymbol{\phi})  }   \right\rangle_T  }.
  \label{tds}
\end{equation}  
The above equation is obtained by imposing the stationarity
condition $\frac{\partial \Phi}{\partial q_{\alpha \beta}}=0$ on $\Phi(\{ q_{\alpha \beta}  \} )$. Combining
Eq. (\ref{rtrt77}) with Eqs. (\ref{uydqs}) and (\ref{fdsa1}), we obtain an expression
for the spectral density in terms of $\Phi(\{ q_{\alpha \beta}  \} )$, that is
\begin{equation}
  \rho_{S^{-1}}(\lambda) = \frac{2}{\pi} \lim_{\epsilon \rightarrow 0^{+}} {\rm Im} \left[ \frac{\partial}{\partial z} \lim_{n \rightarrow 0} \Phi(\{ q_{\alpha \beta}  \} )   \right].
  \label{hdf22}
\end{equation}
Finally, we simplify the saddle-point equations
by making the diagonal replica symmetric {\it ansatz}
\begin{equation}
  q_{\alpha \beta} = \frac{q}{2} \delta_{\alpha \beta} \quad (q \in \mathbb{C}).
  \label{tesc}
\end{equation}  
Inserting the above assumption into Eqs. (\ref{tds}) and  (\ref{hdf22}) and taking the limit $n \rightarrow 0$, we
arrive at the final expression for the spectral density
\begin{equation}
\rho_{S^{-1}}(\lambda) = \frac{1}{\pi} \lim_{\epsilon \rightarrow 0^{+}} {\rm Im} \left\langle \frac{1}{z - \frac{\mu}{T} - g^{2}(T) q }  \right\rangle_T
\end{equation}  
where $q$ solves the fixed-point equation
\begin{equation}
q = \left\langle  \frac{g^{2}(T)}{z - \frac{\mu}{T} - g^{2}(T)q  }  \right\rangle_T.
\end{equation}  
The application of the replica method to determine the spectral density $\rho_{A^{\prime}}(\lambda)$ of the
transformed coupling matrix $\boldsymbol{A}^{\prime}$ unfolds in a similar way. By comparing Eqs. (\ref{adjacency1}) and (\ref{precision}), it is straightforward
to conclude that $\rho_{A^{\prime}}(\lambda)$ is given by
\begin{equation}
\rho_{A^{\prime}}(\lambda) = \frac{1}{\pi} \lim_{\epsilon \rightarrow 0^{+}} {\rm Im} \left\langle \frac{1}{z - \mu - T g^{2}(T) q }  \right\rangle_T,
\end{equation}  
where $q$ fulfills the equation
\begin{equation}
q = \left\langle  \frac{T g^{2}(T)}{z - \mu - T g^{2}(T)q  }  \right\rangle_T.
\end{equation}  

\bibliography{RMT}

\end{document}